\documentclass[pra,twocolumn,showpacs,preprintnumbers,amsmath,amssymb]{revtex4}

\usepackage{graphicx,subfigure}
\usepackage{dcolumn}
\usepackage{bm}

\newcommand{\be}{\begin{equation}}
\newcommand{\ee}{\end{equation}}

\begin{document}

\title{Collisional oscillations of trapped boson-fermion mixtures
approaching collapse}

\author{P. Capuzzi} 
\author{A. Minguzzi} 
\author{M. P. Tosi}
\affiliation{NEST-INFM and
Classe di Scienze, Scuola Normale Superiore, Piazza dei Cavalieri 7,
I-56126 Pisa, Italy}

\date{\today}

\begin{abstract}
We study the collective modes of a confined gaseous cloud of bosons
and fermions with mutual attractive interactions at zero
temperature. The cloud consists of a Bose-Einstein condensate and a
spin-polarized Fermi gas inside a spherical harmonic trap and the
coupling between the two species is varied by increasing either the
magnitude of the interspecies $s$-wave scattering length or the number of
bosons. The mode frequencies are obtained in the collisional regime by
solving the equations of generalized hydrodynamics and are compared
with the spectra calculated in the collisionless regime within a
random-phase approximation. We find that, as the mixture is driven
towards the collapse instability, the frequencies of the modes of
fermionic origin show a blue shift which can become very significant
for large numbers of bosons. Instead the modes of bosonic origin show
a softening, which becomes most pronounced in the very proximity of
collapse. Explicit illustrations of these trends are given for the
monopolar spectra, but similar trends are found for the dipolar and
quadrupolar spectra except for the surface ($n=0$) modes which are
essentially unaffected by the interactions.
\end{abstract}

\pacs{03.75.Kk, 03.75.Ss, 67.60.-g}

\maketitle
\section{\label{sec:intro}Introduction}

Boson-fermion mixtures have been produced with atomic gases in several
experiments \cite{DeMarco1999a,Granade2002a, Truscott2001a,
Schreck2001a,Roati2002a,Hadzibabic2002a} and the deep
quantum-degeneracy regime has been reached for both the bosonic and
the fermionic component \cite{Streckner2003a,Hadzibabic2003a}.  The
aim of fermion cooling in these mixtures has been the realization of a
superfluid state \cite{Stoof1996a,Ohashi2002a}.  Even though the
mixtures are very dilute, the interactions between bosons and fermions
play important roles and in particular the $s$-wave collisions between
the two species are exploited in the evaporative cooling process to
obtain rethermalization. Mean-field interactions affect the expansion
rate of the cloud and its thermodynamic properties
\cite{Roati2002a,Hui2003c}.

For a $^{87}$Rb-$^{40}$K mixture the boson-fermion interactions are
large and attractive \cite{Ferrari2002a}, while in other systems they
could be tuned to become attractive by exploiting Feshbach resonances
\cite{Feshbach1962a,Stwalley1976a,Tiesinga1993a}.  Boson-fermion
attractions enhance the overlap between the two species, favoring in
principle a boson-mediated fermion pairing
\cite{Heiselberg2000a,Bijlsma2000a}, and ultimately lead to the
collapse of the cloud as the boson-fermion coupling strength overcomes
the Fermi pressure. Collapse has been observed experimentally in the
$^{87}$Rb-$^{40}$K mixture as a sudden disappearance of the fermion
cloud when the number of bosons, and hence the coupling between the
two species, is increased \cite{Modugno2002a}.  The equilibrium
properties and the phase diagram of a mixture with attractive
interactions have been investigated by several authors
\cite{Miyakawa2001a,Roth2002a,Roth2002b,Hui2003d}, finding good
agreement between the experimental data and a mean-field theoretical
description \cite{Modugno2003a,Capuzzi2003d}.

The study of the collective modes of a boson-fermion mixture with
repulsive interactions has proved to be a good indicator of the
quantum transition to a spatially demixed state \cite{Capuzzi2003a}.
For the case of attractive interactions some analyses using
semi-analytical methods \cite{Miyakawa2000a,Xia-Ji2003a} have
suggested that the lowest-lying monopolar mode should show a softening
in the approach to collapse. A similar frequency softening is found
for a pure condensate with increasingly large boson attractions
\cite{Dodd1996a,Perez-Garcia1996a}. In contrast, the numerical
solution of the equations of motion for boson-fermion mixtures in the
collisionless Random-Phase-Approximation (RPA) does not give
indications of frequency softening for the choice of parameters
adopted in the calculations \cite{Sogo2002a}. Moreover, it is found
that when the number of bosons is much larger than that of fermions
some families of modes are strongly blue-shifted as a consequence of
the increase of the particle densities in the cloud approaching
collapse \cite{Capuzzi2003b}.  However, the RPA calculations could not
be brought to the very proximity of collapse because of very severe
numerical difficulties.

In this paper we study the collective modes of a trapped boson-fermion
mixture with attractive interactions using the equations of
generalized hydrodynamics. These allow us to span the entire range of
boson-fermion interaction strength up to the collapse point. Although
the hydrodynamic equations are strictly valid only in the collisional
regime, their predictions also serve as a guideline to interpret the
RPA spectra.  By evaluating the frequencies of the first ten low-lying
modes we find that some modes undergo a blue shift which can become
quite large at large numbers of bosons, while some other modes display
a softening which becomes pronounced in a very narrow window in the
proximity to collapse.

\section{\label{sec:model}Equations of generalized hydrodynamics}

We consider a model for a dilute mixture of two species of alkali
atoms inside harmonic traps, one species being spin-polarized fermions
of mass $m_F$ and the other bosons of mass $m_B$. The boson-boson and
boson-fermion interactions are described by contact potentials
involving the coupling constants $g_{BB} = 4\pi\hbar^2\,a_{BB}/m_B$
and $g_{BF}=2\pi\hbar^2 \,a_{BF}/m_r$, where $a_{BB}$ and $a_{BF}$ are
the corresponding $s$-wave scattering lengths and $m_r=m_B
m_F/(m_F+m_B)$. The fermion-fermion interactions are negligible, since
the Pauli principle forbids collisions in the $s$-wave channel between
fermions of the same spin. In the following we treat the case of
boson-fermion attractions ($a_{BF}<0$) and boson-boson repulsions
($a_{BB}>0$).

The dynamics of the mixture in the collisional regime is described by
the equations of generalized hydrodynamics as in Ref.\
\cite{Capuzzi2003a}.  These are (i) the equations for particle
conservation,
\begin{equation}
\partial_t \rho_{\sigma} + \nabla\cdot \mathbf{j}_{\sigma} = 0,
\label{Ec:continuity}
\end{equation}
relating the particle density $\rho_{\sigma}(\mathbf{r},t)$ to the
current density $\mathbf{j}_{\sigma}(\mathbf{r},t)$ for each component
($\sigma=B,F$); and (ii) the quantum Navier-Stokes equations for
momentum conservation, which are
\begin{equation}
m_{\sigma}\,\partial_t \mathbf{j}_{\sigma} = \rho_{\sigma}\,
(\mathbf{F}_{\sigma} - \nabla V_{\sigma} - g_{BF}\,\nabla
\rho_{\bar{\sigma}})
\label{Ec:mom}
\end{equation}
to linear order in the velocity fields. Here $\bar{\sigma}$ denotes
the component different from $\sigma$,
$V_{\sigma}(\mathbf{r})=m_{\sigma}\omega_{\sigma}^2\,r^2/2$ are
isotropic trapping potentials, and $\mathbf{F}_{\sigma}(\mathbf{r},t)$
are self-consistent internal forces. We take for the latter the
expressions
\begin{equation}
\mathbf{F}_B = -\nabla\left[ g_{BB}\,\rho_B - \frac{\hbar^2}{2\,m_B}\, 
\frac{\nabla^2\sqrt{\rho_B}}{\sqrt{\rho_B}}\right]
\label{Ec:FB}
\end{equation}
and 
\begin{equation}
\mathbf{F}_F = -\nabla \left[ A\,\rho_F^{2/3} -
  \frac{\hbar^2}{6\,m_F}\,
  \frac{\nabla^2\sqrt{\rho_F}}{\sqrt{\rho_F}}\right] , 
\label{Ec:FF}
\end{equation}
with $A=\hbar^2\,(6\pi^2)^{2/3}/2m_F$.  The equations of generalized
hydrodynamics are thereby closed by means of a local-density
approximation on the kinetic stress tensors, which also includes
surface kinetic contributions through the last term on the rhs of
Eqs.~(\ref{Ec:FB}) and (\ref{Ec:FF}) \cite{Capuzzi2003a}.  These
surface correction terms transcend the standard Thomas-Fermi
approximation and for bosons the present approach is in fact
equivalent to the full time-dependent Gross-Pitaevskii equation.

The equilibrium state of the mixture is given by the stationary
solutions of Eq.\ (\ref{Ec:mom}). These also correspond to a minimum
of the energy functional
\begin{eqnarray}
E[\rho_B, \rho_F] &=& \int d^3r\left(V_B\,\rho_B + \frac{g_{BB}}{2}
\rho_B^2 + \xi_B\right) \nonumber \\ && + \int d^3r \left( V_F\,\rho_F
+ \frac{3}{5}\,A\,\rho_F^{5/3} + \xi_F\right) \nonumber \\ &&+
g_{BF}\int d^3r \rho_F\,\rho_B,
\label{Ec:Energy}
\end{eqnarray} 
where the surface kinetic-energy terms are 
$\xi_B
= \hbar^2|\nabla\sqrt{\rho_B}|^2/2m_B $ and 
$\xi_F =
\hbar^2|\nabla\sqrt{\rho_F}|^2/6m_F $.

At increasingly large boson-fermion attractions the densities of the
two species increase in their overlap region and collapse occurs when
these attractions overcome the Fermi kinetic pressure and the
boson-boson repulsions. Collapse is identified in the numerical search
for the equilibrium state as the point where it is no longer possible
to find a stable minimum for the energy functional in Eq.\ (\ref{Ec:Energy}).
We have verified that the location of collapse found in this way is
well approximated by the estimate \cite{Molmer98a,Miyakawa2001a}
\begin{equation}
|a_{BF}^{c}| = \left(\frac{a_{BB}}{\alpha\,k_F}\right)^{1/2},
\end{equation}
where $k_F=(6\pi^2\,\rho_F(0))^{1/3}$ is the Fermi wave number
estimated from the fermion density at the
center of the trap and $\alpha = [3^{1/3}/ (2\pi)^{2/3}]
(m_F+m_B)^2/(4m_Fm_B)$.

In the following we characterize the approach of the mixture to the
 collapse instability by following the behavior of its low-lying
 collective oscillation frequencies.  The normal modes of
 Eqs.~(\ref{Ec:continuity}) and (\ref{Ec:mom}) are found by expanding
 $\rho_{\sigma}(\mathbf{r},t)$ around the equilibrium state
 $\rho_{\sigma}^0(r)$ as $\rho_{\sigma} (\mathbf{r},t) =
 \rho_{\sigma}^{0}(r) + \delta \rho_{\sigma}
 (\mathbf{r})\,e^{i\,\omega\,t}$, linearizing Eqs.\
 (\ref{Ec:continuity}) and (\ref{Ec:mom}), and Fourier transforming with
 respect to the time variable. This  yields the coupled eigenvalue equations
\begin{equation} 
m_{\sigma}\,\omega^2\delta\rho_{\sigma} = \nabla\cdot(\rho_{\sigma}^0\,
\delta\mathbf{F}_{\sigma}) - g_{BF}\,\nabla\cdot(\rho_{\sigma}^0 \nabla
\delta\rho_{\bar{\sigma}}), 
\label{Ec:collmodes}
\end{equation}
where $\delta\mathbf{F}_{\sigma}$ are the forces obtained as
$\delta\mathbf{F}_{\sigma} =
\mathbf{F}_{\sigma}[\rho_{\sigma}^0+\delta\rho_{\sigma}]-
\mathbf{F}_{\sigma}[\rho_{\sigma}^0]$ to linear order in the density
fluctuations (for their expressions see Ref.\ \cite{Capuzzi2003a}).
We also determine the solutions of the eigenvalue equations in the
case where the dynamical coupling between the two components is
neglected.  Hereafter we shall call these solutions the uncoupled
boson and fermion modes and calculate them by solving Eq.\
(\ref{Ec:collmodes}) for $\sigma=B$ and $F$ with the second term in
the rhs set to zero. This calculation will help us in attributing
bosonic or fermionic character to the coupled modes of the mixture by
comparison with the frequencies of the uncoupled modes. Even though
the bosonic and fermionic oscillations that are obtained in this way
are dynamically uncoupled, the effect of the mutual boson-fermion
interaction is included in their calculation through the use of the
density profiles of the mixture at equilibrium.

The numerical procedure that we have used can be summarized as
follows: (i) we find the equilibrium density profiles by minimizing
the energy functional (\ref{Ec:Energy}) using the steepest-descent
method; (ii) we project Eq.\ (\ref{Ec:collmodes}) into subspaces of
different angular momentum $l$ by factorizing the density fluctuations
as $\delta\rho_{\sigma}(\mathbf{r}) = \delta\rho_{\sigma}^{l}(r) \,
Y_{lm}(\hat{r})$ where $Y_{lm}(\hat{r})$ are the spherical harmonics;
and (iii) we solve the coupled equations in a given $l$-subspace by
means of standard linear-algebra routines \cite{lapack3}.

\section{\label{sec:modes}Collective modes}

We have solved Eq.\ (\ref{Ec:collmodes}) for two sets of
experimentally relevant system parameters, corresponding to a
$^7$Li-$^6$Li mixture \cite{Schreck2001a} and to a $^{87}$Rb-$^{40}$K
mixture \cite{Modugno2002a}.  In both cases we have set the frequency
of the isotropic trap at the geometric average of the frequencies in
the experimental setup and have adopted the experimental value of the
boson-boson scattering length. In our calculations we follow two
different routes to reach collapse.  We drive the $^7$Li-$^6$Li
mixture to collapse by varying the value of $a_{BF}$ from positive to
strongly negative, as could be achieved by means of a Feshbach
resonance.  In the $^{87}$Rb-$^{40}$K mixture instead we keep $a_{BF}$
fixed and increase the number $N_B$ of bosons, as is done in the
experiments of Modugno {\it et al.}  \cite{Modugno2002a}.

\subsection{The $^7$Li-$^6$Li mixture} 

Figure~\ref{fig:collLiLi3} shows the frequencies of the low-lying
monopole ($l=0$) modes in the $^7$Li-$^6$Li mixture as functions of
$a_{BF}$, with the choice of parameters $\omega_F=\omega_B=2\pi\times
1000\,{\rm s}^{-1}$, $a_{BB}=0.27\,$nm, and particle numbers
$N_F=10^4$ and $N_B=10^6$.  Pronounced blue shifts are observed at
intermediate values of the boson-fermion scattering length both in a
set of eigenfrequencies of Eq.~(\ref{Ec:collmodes}) (circles) and in
the frequencies of the uncoupled fermionic modes (dashed lines).  This
effect is a consequence of the increase of fermion density in the
central part of the trap as the strength of the coupling is
increased. For our choice of parameters the bosonic density profile is
almost unaffected in this range of boson-fermion coupling and
essentially acts as an effective attractive well for the fermions. The
proximity of these modes to the uncoupled fermionic modes indicates
that the effect of the dynamical coupling is negligible, thus
confirming the interpretation in terms of static mean-field effects.
At larger values of $|a_{BF}|$ the other eigenfrequencies of Eq.\
(\ref{Ec:collmodes}) (circles) show an increasing departure from the
uncoupled bosonic modes (solid lines) and tend to rapidly soften as
collapse is approached. We conclude that, in contrast to the
blue-shift of the fermionic modes, the softening of ``bosonic'' modes
is a truly dynamical signature of the impending collapse.

Similar trends of the mode frequencies are also found for the dipole
($l=1$) and quadrupole ($l=2$) oscillations, except that the lowest
($n=0$) mode is in both cases essentially unaffected by the
interactions. For the dipole mode this behavior is an exact
consequence of the generalized Kohn theorem
\cite{Kohn1961a,Dobson1994a}, while for modes of higher $l$ it is
strictly valid only in the Thomas-Fermi limit. This can be explicitly
verified from Eq.~(\ref{Ec:collmodes}) by making the Ansatz
$\delta\rho_{\sigma}(\mathbf{r}) \propto r^l\,Y_{lm}(\hat{r})$ for the
density fluctuations and by exploiting the property
$\nabla^2\delta\rho_{\sigma}=0$ for these modes.

\subsection{The $^{87}$Rb-$^{40}$K mixture} 

We evaluate the collective modes of a $^{87}$Rb-$^{40}$K mixture with
scattering lengths $a_{BB}=5.5\,$nm and $a_{BF}=-21.7\,$nm.  Although
both components are inside the same magnetic trap, their trapping
frequencies differ considerably as a consequence of the large
difference in atomic masses. In the numerical calculations we take
$\omega_B = 2\pi\times 90.9\,{\rm s}^{-1}$ and $\omega_F = 2\pi\times
134\,{\rm s}^{-1}$.  Collapse is approached by varying $N_B$ from
$4\times 10^3$ to approximately $8\times 10^4$, while keeping
$N_F=2\times 10^4$ fixed.

\begin{figure}
\centering
\includegraphics[width=0.9\columnwidth]{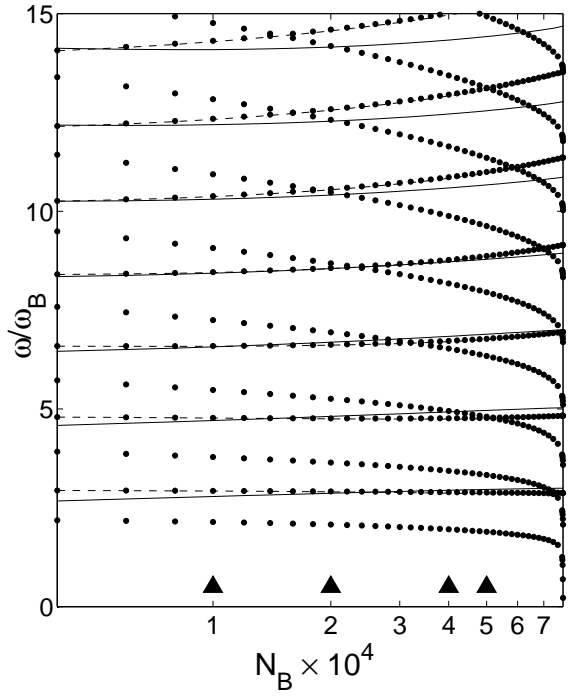}
\caption{\label{fig:collRbK2} Monopolar frequencies $\omega$ (in units
  of $\omega_B$) of the collisional collective modes as functions of
  the number $N_B$ of bosons (in log scale) for the $^{87}$Rb-$^{40}$K
  mixture with $N_F=2\times 10^4$.  The solid and dashed lines show
  the frequencies of the dynamically uncoupled modes for bosons and
  fermions, respectively. The triangles correspond to the values of
  $N_B$ for which we have performed the RPA calculations of Fig.\
  \ref{fig:comparRK}.}
\end{figure}

In Fig.\ \ref{fig:collRbK2} we show the frequencies of the low-lying
monopolar modes for this mixture. At increasing values of the boson
number we find that the frequencies of the bosonic modes, identified
as those that are closer to the uncoupled bosonic fluctuations, show a
monotonic decrease which becomes very pronounced very close to
collapse.  On the other hand, the fermionic modes only show a modest
blue shift. This is not as large as in our results for the
$^7$Li-$^6$Li mixture, since the numbers of particles of the two
species are here more similar.

\section{Spectral functions}

We have also calculated the whole spectral functions under the effect
of the perturbing fields $\delta U_{\sigma}(\mathbf{r})e^{i\omega t}$,
in order to make a comparison of the collisional spectra with those of
the mixtures in the RPA collisionless regime. Peaks in the spectral
functions indicate the location of collective modes, while their
height measures the probability of a transition from the equilibrium
state to an excited state with energy $\hbar\omega$.

In the collisional regime the spectral peaks are located in
correspondence of the eigenfrequencies $\omega_i$ of Eq.\
(\ref{Ec:collmodes}), while the strength of a transition at a
frequency $\omega$ is estimated from the corresponding density
fluctuations $\delta\rho_{\sigma}^i(\mathbf{r})$ as 
\begin{equation} 
S_{\sigma\sigma'}^{\text{coll}}(\omega) = -\frac{1}{\pi}{\rm
Im}\sum_{i} \frac{\mathcal{Z}_{\sigma'}^i}{\omega_i^2 -
\omega^2} \int d^3r \,\delta
U_{\sigma}^{*}(\mathbf{r})\,\delta\rho_{\sigma'}^i(\mathbf{r})
\label{Eq:Scoll}
\end{equation}
(see Appendix).  Here the sum is performed over the whole spectrum of
Eq.\ (\ref{Ec:collmodes}) and $\mathcal{Z}_{\sigma}^i$ is given by
\begin{equation}
\mathcal{Z}_{F}^i = \frac{2\,A}{3\,m_F}
\int d^3r\,{(\rho_F^0)}^{-1/3}\,\delta\rho_F^i\,
\nabla\cdot\left(\rho_F^0\nabla\delta U_F\right)
\label{Ec:ZfTF}
\end{equation}
and
\begin{equation}
\mathcal{Z}_B^i = \frac{g_{BB}}{m_B} \int d^3r\,
\delta\rho_B^i\,\nabla\cdot\left(\rho_B^0\nabla\delta U_B\right).
\label{Ec:ZbTF}
\end{equation}
Since we are using in Eq.~(\ref{Eq:Scoll}) the equilibrium profiles
and the mode frequencies of the fully coupled dynamical equations, we
expect it to yield the spectral functions quite accurately.

More generally, the spectral functions are related to the
density-density responses
$\chi_{\sigma\sigma'}(\mathbf{r},\mathbf{r}',\omega)$ by the
fluctuation-dissipation theorem,
\begin{equation}
S_{\sigma\sigma'}(\omega) = -\frac{1}{\pi}{\rm Im}\int d^3r\,d^3r'\,\delta
U_{\sigma}^{*}(\mathbf{r}) \chi_{\sigma\sigma'}
(\mathbf{r},\mathbf{r}',\omega)\,\delta U_{\sigma'}(\mathbf{r}').
\label{Ec:responses}
\end{equation}  
We evaluate the response functions in the collisionless regime within
the RPA, following the approach described in
Refs. \cite{Capuzzi2001a,Capuzzi2003b}.  For a monopolar drive
($\delta U_{\sigma} ({\mathbf r})\propto r^2$) we calculate both the
fermionic response $S_{FF}^{l=0}(\omega)$ and the bosonic response
$S_{BB}^{l=0}(\omega)$, corresponding to excitations induced by
applying the drive separately on the fermions or on the bosons.

\begin{figure}
\centering
\includegraphics[width=0.9\columnwidth]{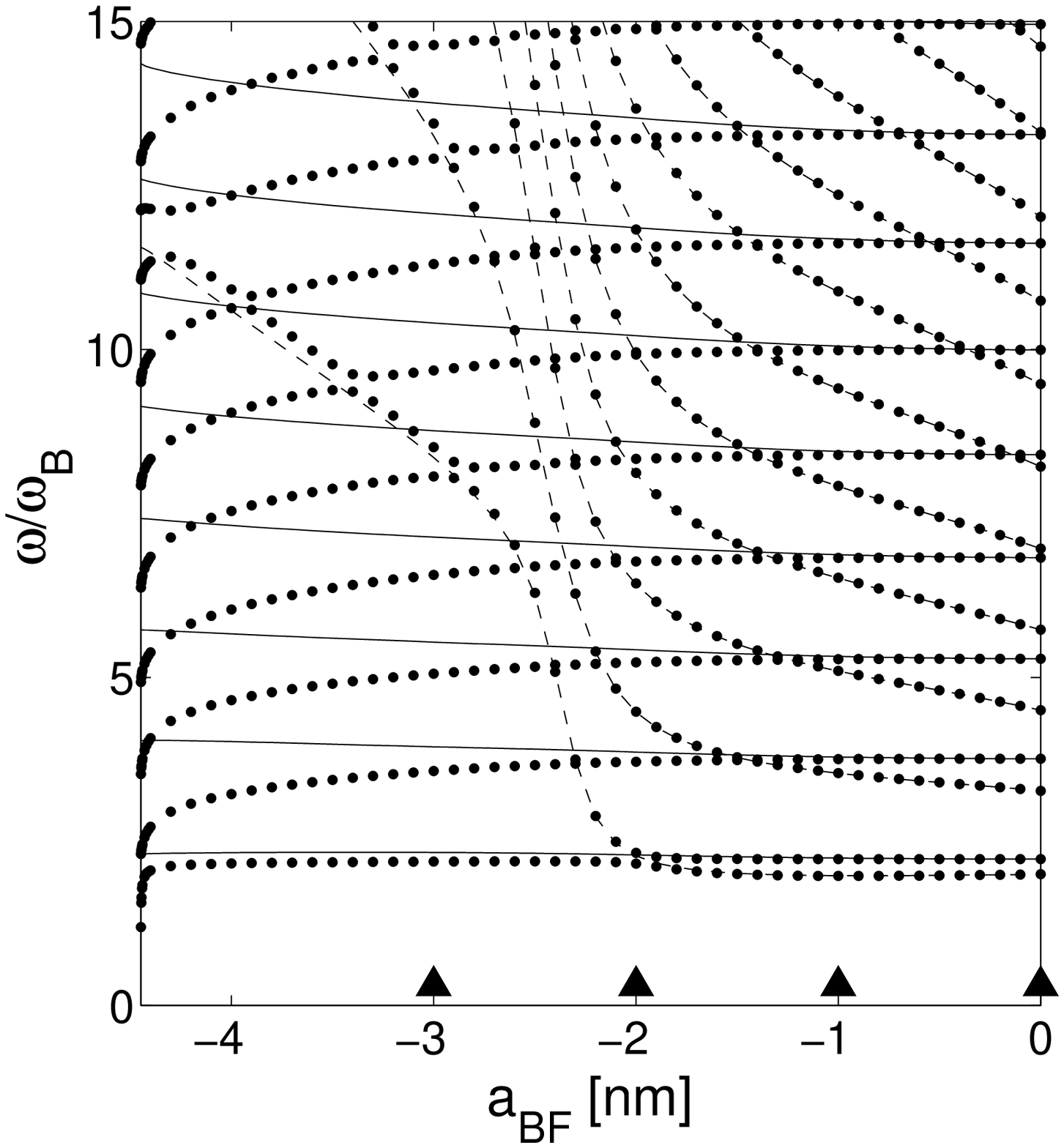}
\caption{\label{fig:collLiLi3} Monopolar frequencies $\omega$ (in
 units of the bare boson-trap frequency $\omega_B$) of the collisional
 collective modes as functions of the mutual scattering length
 $a_{BF}$ (in nm) for a $^6$Li-$^7$Li mixture with $N_F=10^4$ and
 $N_B=10^6$.  The solid and dashed lines show the frequencies of the
 dynamically uncoupled modes for bosons and fermions,
 respectively. The triangles correspond to the values of $a_{BF}$ for
 which we have performed the RPA calculations of Figs.\
 \ref{fig:comparFF} and \ref{fig:comparBB}. }
\end{figure}

\begin{figure}
\includegraphics[width=\linewidth]{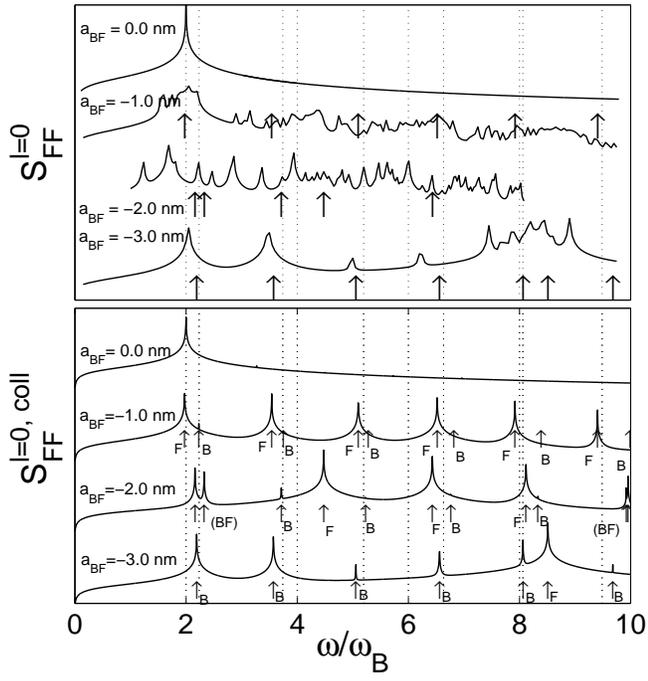}
\caption{\label{fig:comparFF} Monopolar spectrum (in log scale and
  arbitrary units) for fermions as functions of $\omega$ (in units of
  $\omega_B$) in a $^6$Li-$^7$Li mixture with $N_F=10^4$, $N_B=10^6$
  and for various values of $a_{BF}$. The spectra are plotted for the
  sake of clarity with a spectral width of order $10^{-5}\,\omega_B$.
  The vertical dotted lines indicate the values of the bosonic and
  fermionic monopolar modes in the absence of boson-fermion
  coupling. Top panel: spectral response within the RPA; the arrows
  indicate the frequencies of the collisional modes with
  non-negligible spectral weight for the same parameters. Bottom
  panel: spectral response in the collisional regime; the arrows
  denote all collisional modes for the same parameters with label B or
  F to indicate the type of mode as identified from its proximity to
  an uncoupled collisional mode (the labels BF indicate ambiguous
  cases).}
\end{figure}

\begin{figure}
\includegraphics[width=\linewidth]{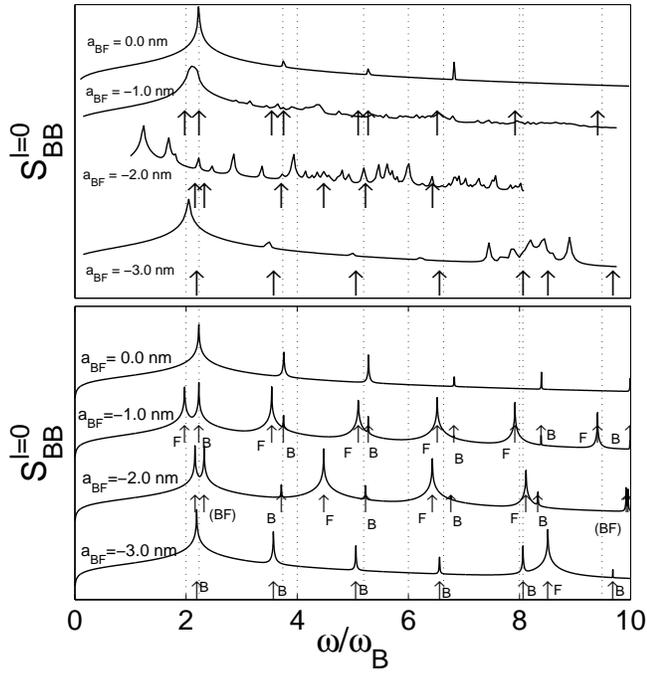}
\caption{\label{fig:comparBB} The same as in Fig.\ \ref{fig:comparFF}
for the bosonic response.}
\end{figure}

In Figs.\ \ref{fig:comparFF} and \ref{fig:comparBB} we report the
spectral functions of the $^7$Li-$^6$Li mixture as obtained from the
RPA formalism (top panels) and from Eq.\ (\ref{Eq:Scoll}) for the
collisional regime (bottom panels) at various values of $a_{BF}$ and
with the same choice of parameters as in Fig.~\ref{fig:collLiLi3}.  In
the absence of boson-fermion coupling a fermionic perturbing field
excites the pure $n=0$ mode of the fermions, which gives the only
contribution to the FF spectra. In contrast, the BB spectra contain in
the same limit several weak peaks at higher frequencies above the pure
$n=0$ mode of the bosons, owing to the boson-boson interactions. For
$a_{BF}\neq 0$ the RPA spectra are very complex, from the lifting of
degeneracy in the discrete-level structure which is due to the
interactions with the bosons. The fermionic peaks can be recognized as
broad fragmented contributions of largest height in the fermionic
spectral functions, while the bosonic peaks are usually isolated and
are most marked in the bosonic spectra \cite{Capuzzi2003b}.

Quite interestingly, the spectra in the two regimes display similar
gross features.  At small values of $|a_{BF}|$ the peaks are located
near the frequencies of the higher-order monopolar modes, but at
larger values they tend to spread over the whole range of
frequency. This phenomenon occurs for the same choice of parameters
for which the blue shift of the collisional fermionic modes becomes
pronounced and several mode crossings occur.  For $a_{BF}=-3$ nm the
first fermionic peak is strongly blue-shifted while the low-frequency
part of the spectrum is dominated by bosonic modes having considerable
oscillator strength in both bosonic and fermionic spectral
functions. This indicates a strong dynamical coupling of the density
fluctuations of the two species in a regime of parameters where the
uncoupled frequencies are close to the coupled solutions for the
collisional fermionic modes (see again Fig.~\ref{fig:collLiLi3}).  A
careful analysis also shows that for $a_{BF}=-3$ nm the frequencies of
the bosonic modes in the RPA spectra are slightly red-shifted, showing
the first indications of a trend towards softening on the approach to
collapse.

\begin{figure}
\includegraphics[width=\columnwidth]{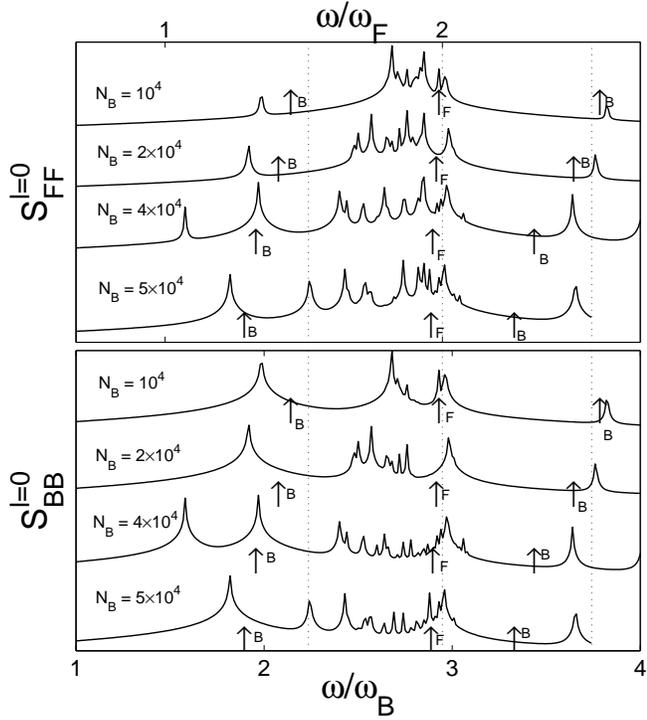}
\caption{\label{fig:comparRK} Monopolar spectral response (in log
  scale and arbitrary units) from the RPA for fermions (top) and
  bosons (bottom) as functions of $\omega$ (in units of $\omega_F$ or
  $\omega_B$) for a $^{87}$Rb-$^{40}$K mixture with $a_{BB}=5.5\,$nm,
  $a_{BF}=-21.7\,$nm, $N_F=2 \times 10^4$, and increasing values of
  $N_B$.  The arrows indicate the collisional modes for the same
  parameters. The vertical dotted lines are the values of the bosonic
  and fermionic monopolar modes in the absence of boson-fermion
  coupling.}
\end{figure}

Finally, in Fig.\ \ref{fig:comparRK} we compare the collisional
results for the monopolar spectral function (bottom panel) with those
of the RPA (top panel) for a $^{87}$Rb-$^{40}$K mixture with the same
choice of parameters as in Fig.~\ref{fig:collRbK2}. The RPA spectra
show a fragmented fermionic contribution around the frequency
$2\omega_F$ of the bare monopolar oscillation as well as two main
bosonic peaks at its sides. The frequency of these two peaks tends to
decrease with increasing $N_B$, similarly to what is found for the
same modes in the collisional regime.  This suggests that such
frequency softening is a signature of the incipient collapse.

\section{\label{sec:conclu}Summary and concluding remarks}

In summary, we have used the equations of generalized hydrodynamics to
study the collective modes of trapped boson-fermion mixtures with
mutual attractive interactions driving the mixture towards the
collapse instability.  We have focused on two specific systems of
experimental interest, for which we have chosen two different routes
to collapse.

In both cases we find that the collective spectra show a frequency
softening of a family of modes of bosonic nature as a signature of the
incipient collapse. This softening becomes most pronounced in a very
narrow region of parameters near collapse.  A second effect of
increasing the attractive coupling between the two species is a blue
shift of the modes of fermionic origin, which becomes more evident for
large numbers of bosons and reflects the compression of the fermionic
cloud from the interactions with the bosons.

A comparison of the hydrodynamic spectra with the spectra calculated
in the collisionless regime within the random-phase approximation
suggests that both the blue shift of the ``fermionic'' modes and the
softening of the ``bosonic'' modes are general features of the
dynamics of the mixture. By comparing the height of the peaks in the
bosonic and fermionic spectral functions we also conclude that the two
families of modes can be most efficiently excited by applying the
driving fields separately on the species.

Our analysis has been restricted to the linear regime, neglecting
nonlinear and beyond-mean-field effects which may start to play a role
close to collapse. We hope to address these issues in future work.

\begin{acknowledgments}
This work was partially supported by INFM through the PRA-Photonmatter
Program.
\end{acknowledgments}

\appendix*
\section{Spectral functions in the collisional regime}
The spectral functions of a mixture in the collisional regime can be
obtained by adding the perturbing fields $\delta
U_{\sigma}(\mathbf{r})e^{i\omega t}$ to the trapping potential
$V_{\sigma}(\mathbf{r})$ in Eqs.\ (\ref{Ec:continuity}) and
(\ref{Ec:mom}) and then linearizing the equations of motion in terms
of the induced density fluctuation $\delta
\rho_{\sigma}(\mathbf{r},\omega)$. This leads in Fourier transform
with respect to time to the equation
\begin{equation}
m_{\sigma}\,\omega^2\delta\rho_{\sigma} = \nabla\cdot(\rho_{\sigma}^0
\,\delta\mathbf{F}_{\sigma})-g_{BF}\,\nabla\cdot(\rho_{\sigma}^0
\nabla\delta\rho_{\bar{\sigma}}) - \nabla\cdot(\rho_{\sigma}^0
\nabla\delta U_{\sigma})
\label{Ec:hydrocoll}
\end{equation}
for each component of the mixture.  Here, at variance from
Eq.~(\ref{Ec:collmodes}) $\omega$ is fixed by the frequency of the
drive.  Equation (\ref{Ec:hydrocoll}) can be solved by expanding
$\delta\rho_{\sigma}(\mathbf{r},\omega)$ in the basis of the
eigenvectors $\delta\rho_{\sigma}^i(\mathbf{r})$ of
Eq.~(\ref{Ec:collmodes}) as
\begin{equation}
\delta\rho_{\sigma}(\mathbf{r},\omega) = \sum_{i}C^i(\omega)
\,\delta\rho_{\sigma}^i(\mathbf{r}).
\label{Eq:drhocoll}
\end{equation}
In turn, the expansion coefficients $C^i$ are obtained by projecting
Eq.\ (\ref{Ec:hydrocoll}) into the subspace generated by a given
fluctuation $\delta\rho_{\sigma}^j(\mathbf{r})$. This procedure
requires defining a scalar product between two eigenmodes of Eq.\
(\ref{Ec:collmodes}), which we shall denote by the braket $\langle
\delta\rho^i | \delta\rho^j\rangle $.  The projection yields
\begin{equation}
C^i(\omega) = \frac{\langle
\delta\rho^i | \nabla \cdot\left(\rho_{\sigma}^0\,
\nabla\delta U_{\sigma}\right)\rangle} {\omega_i^2 -
\omega^2}
\end{equation} 
where $\omega_i$ is the frequency corresponding to the eigenvector
$\delta\rho_{\sigma}^i(\mathbf{r})$.

A generic scalar product can be written as
\begin{equation}
\langle \delta\rho^i|\delta\rho^j\rangle = \sum_{\sigma,\,\sigma'}
\int d^3r\,\delta\rho_{\sigma}^i(\mathbf{r})\,w_{\sigma\sigma'}(\mathbf{r})\,
\delta\rho_{\sigma'}^j(\mathbf{r})
\label{Ec:scalar}
\end{equation}
where $w_{\sigma\sigma'}(\mathbf{r})$ are suitable weights to be
determined.  Here we take for the scalar product (\ref{Ec:scalar})
that of a bosonic and of a fermionic cloud in the Thomas-Fermi limit
with vanishing mutual interactions. We thus set $w_{FF} =
2A(\rho_F^0)^{-1/3}/3m_F$ and $w_{BB}=g_{BB}/m_B$ and neglect the
off-diagonal weight functions. This choice yields
\begin{equation}
C^i_{\sigma}(\omega) = \frac{\mathcal{Z}_{\sigma}^i}{\omega_i^2 - \omega^2}
\label{Eq:Cdef}
\end{equation}
for fermions and bosons, where $\mathcal{Z}_{\sigma}^i$ is given in
Eqs.\ (\ref{Ec:ZfTF}) and (\ref{Ec:ZbTF}).  We have taken
$\delta\rho_{\sigma}^i$ as normalized to unity ( $\langle
\delta\rho^i| \delta \rho^i \rangle = 1$).

The spectral functions are then calculated from the definition
\begin{equation}
S_{\sigma\sigma'}(\omega) = -\frac{1}{\pi}{\rm Im}\int d^3r\,\delta
U_{\sigma}^{*}(\mathbf{r})\,\delta\rho_{\sigma'}(\mathbf{r}, \omega)
\label{Eq:Scollapp}
\end{equation}
where $\delta\rho_{\sigma'}$ is evaluated by setting $\delta
U_{\bar{\sigma}'}=0$.  The use of Eqs.\ (\ref{Eq:drhocoll}) and
(\ref{Eq:Cdef}) in Eq.\ (\ref{Eq:Scollapp}) directly leads to Eq.\
(\ref{Eq:Scoll}) in the main text.


\end{document}